\documentclass[twocolumn,aps,prb,floats,showpacs]{revtex4}

\usepackage{epsf}

\newcommand{\bleq}{\begin{widetext}}
\newcommand{\eleq}{\end{widetext}}

\begin{document}
\draft
\title{Voltage-probe and imaginary potential models for dephasing in a chaotic quantum dot}

\author{P. W. Brouwer and C. W. J. Beenakker}
\address{Instituut-Lorentz, University of Leiden, P.O. Box 9506, 2300 RA
Leiden, The Netherlands}

\begin{abstract}
We compare two widely used models for dephasing in a chaotic quantum dot: The introduction of a fictitious voltage probe into the scattering matrix and the addition of an imaginary potential to the Hamiltonian. We identify the limit in which the two models are equivalent and compute the distribution of the conductance in that limit. Our analysis explains why previous treatments of dephasing gave different results. The distribution remains non-Gaussian for strong dephasing if the coupling of the quantum dot to the electron reservoirs is via ballistic single-mode point contacts, but becomes Gaussian if the coupling is via tunneling contacts.
\smallskip\\
\pacs{PACS numbers: 05.45.+b, 72.10.Bg, 73.40.Gk, 85.30.Vw}
\end{abstract}

\maketitle


\section{Introduction}

Extensive theoretical work has provided a detailed description of the universal features of phase-coherent transport in classically chaotic systems, such as universal conductance fluctuations, weak localization, and a non-Gaussian conductance distribution.\cite{JBS,Zirnbauer,PrigodinEfetovIida,BarangerMello,JPB,BrouwerBeenakker1994,BarangerMello1995,BrouwerBeenakker1995,Efetov1995,AL,McCannLerner,BeenakkerReview} The advances of submicron technology in the past decade have made these manifestations of quantum chaos in electronic transport accessible to experiment.\cite{Marcus1992,Marcus1993,Prober,Chang1994,Bird1994,Bird1995,Chan,Clarke} Although experiments on semiconductor quantum dots confirm the qualitative predictions of the phase-coherent theory, a quantitative comparison requires that loss of phase coherence be included into the theory. Two methods have been used for this purpose.

The first method, originating from B\"uttiker,\cite{Buettiker} is to include a fictitious voltage probe into the scattering matrix. The voltage probe breaks phase coherence by removing electrons from the phase-coherent motion in the quantum dot, and subsequently reinjecting them without any phase relationship. The conductance $G_{\phi}$ of the voltage probe (in units of $2e^2/h$) is set by the mean level spacing $\Delta$ in the quantum dot and the dephasing time $\tau_{\phi}$, according to $G_{\phi} = 2 \pi \hbar/\tau_{\phi} \Delta$. This method was used in Refs.\ \onlinecite{BarangerMello1995}, \onlinecite{BrouwerBeenakker1995}, \onlinecite{Marcus1992}, and \onlinecite{Clarke}. The second method is to include a (spatially uniform) imaginary potential in the Hamiltonian, equal to $-i \hbar/2 \tau_{\phi}$. This method was used in Refs.\ \onlinecite{Efetov1995} and \onlinecite{McCannLerner}.

The two methods have given very different results for the distribution of the conductance $G$, in particular in the case that the current through the quantum dot flows through single-mode point contacts. While the distribution $P(G)$ becomes a delta peak at the classical conductance for very strong dephasing ($\tau_{\phi} \to 0$) in the voltage-probe model, $P(G)$ peaks at zero conductance in the imaginary potential model. It is the purpose of the present paper to reconcile the two methods, and to compute the conductance distribution in the limit that the two methods are equivalent.

The origin of the differences lies with certain shortcomings of each model. On the one hand, the imaginary potential model does not conserve the number of electrons. We will show how to correct for this, thereby resolving an ambiguity in the formulation of the model noted by McCann and Lerner.\cite{McCannLerner} On the other hand, the voltage-probe model describes spatially localized instead of spatially uniform dephasing. This is perfectly reasonable for dephasing by a real voltage probe, but it is not satisfactory if one wants a fictitious voltage probe to serve as a model for dephasing by inelastic processes occurring uniformly in space. A second deficiency of the voltage-probe model is that inelastic scattering requires a continuous tuning parameter $\tau_{\phi}$, while the number of modes $N_{\phi}$ in the voltage probe can take on integer values only. Although the introduction of a tunnel barrier (transparency $\Gamma_{\phi}$) in the voltage probe allows the conductance $G_{\phi} = N_{\phi} \Gamma_{\phi}$ to interpolate between integer values, the presence of {\em two} model parameters creates an ambiguity: The conductance distribution depends on $N_{\phi}$ and $\Gamma_{\phi}$ separately, and not just on the product $N_{\phi} \Gamma_{\phi}$ set by the dephasing time.

In this paper we present a version of the voltage-probe model that does not suffer from this ambiguity and that can be applied to dephasing processes occurring uniformly in space. This version is equivalent to a particle-conserving imaginary potential model. We show that the absorbing term in the Hamiltonian can be replaced by an absorbing lead (the voltage probe) in the limit $N_{\phi} \to \infty$, $\Gamma_{\phi} \to 0$ at fixed $G_{\phi} = N_{\phi} \Gamma_{\phi}$. This is the ``locally weak absorption limit'' of Zirnbauer.\cite{Zirnbauer} Both shortcomings of the voltage-probe model are cured: The limit $N_{\phi} \to \infty$ together with ergodicity ensures spatial uniformity of the dephasing, while the conductance $G_{\phi}$ is the only variable left to parameterize the dephasing rate.
 
The outline of the paper is as follows. In Sec.\ \ref{sec:2} we recall the voltage-probe model and derive the limit $N_{\phi} \to \infty$, $\Gamma_{\phi} \to 0$ at fixed $N_{\phi} \Gamma_{\phi}$ from the particle-conserving imaginary potential model. We then calculate the effect of dephasing on the conductance distribution in the case of single-mode point contacts (Sec.\ \ref{sec:3}). The distribution narrows around the classical series conductance of the two point contacts when the dimensionless dephasing rate $\gamma = 2 \pi \hbar/\tau_{\phi} \Delta$ becomes $\gg 1$, but not precisely in the way which was computed in Refs.\ \onlinecite{BarangerMello1995} and \onlinecite{BrouwerBeenakker1995}. In Sec.\ \ref{sec:4} we briefly consider the case of multi-mode point contacts (number of modes $\gg 1$), which is less interesting. We conclude in Sec.\ \ref{sec:5}.

\section{Two models for dephasing} \label{sec:2}

\begin{figure}
\hspace{0.15\hsize}
\epsfxsize=0.6\hsize
\epsffile{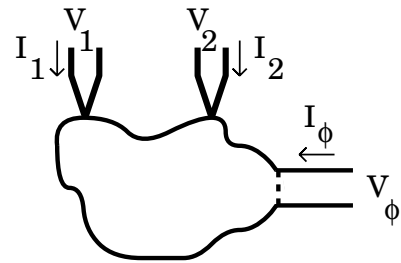}
\vspace{0.3cm}
\caption{\label{fig:1}
Chaotic cavity, connected to current source and drain reservoirs (1 and 2), and to a voltage probe ($\phi$). The voltage probe contains a tunnel barrier (dotted line). The voltage $V_{\phi}$ is adjusted such that $I_{\phi} = 0$.}
\end{figure}

The system under consideration is shown in Fig.\ \ref{fig:1}. It consists of a chaotic cavity, coupled by two point contacts (with $N_1$ and $N_2$ propagating modes at the Fermi energy $E_{F}$) to source and drain reservoirs at voltages $V_1$ and $V_2$. A current $I = I_1 = -I_2$ flows from source to drain. In the voltage-probe model,\cite{Buettiker} a fictitious third lead ($N_{\phi}$ modes) connects the cavity to a reservoir at voltage $V_{\phi}$. Particle conservation is enfor\-ced by adjusting $V_{\phi}$ in such a way that no current is drawn ($I_{\phi} = 0$). The third lead contains a tunnel barrier, with a transmission probability $\Gamma_{\phi}$ which we assume to be the same for each mode. The scattering matrix $S$ has dimension $M = N_1 + N_2 + N_{\phi}$ and can be written as
\begin{equation}
  S = \left( \begin{array}{ccc} s_{11} & s_{12} & s_{1{\phi}} \\
                      s_{21} & s_{22} & s_{2{\phi}} \\
                      s_{{\phi}1} & s_{{\phi}2} & s_{{\phi}{\phi}} \end{array} \right), \label{eq:Sdecomp}
\end{equation}
in terms of $N_i \times N_j$ reflection and transmission matrices $s_{ij}$. Application of the relations\cite{Buettiker2}
\begin{mathletters}  \label{Buetteq}
\begin{eqnarray}
  I_{k\hphantom{l}} &=& {2e^2 \over h} \sum_{l} G_{kl} V_{l}, \ \  k = 1,2,{\phi}, \\
  G_{kl} &=& \delta_{kl} N_k - \mbox{tr}\, s_{kl}^{\phantom \dagger} s_{kl}^{\dagger},
\end{eqnarray}
\end{mathletters}%
yields the (dimensionless) conductance $G = (h/2e^2)I/(V_1 - V_2)$,
\begin{eqnarray}
  G &=& - G_{12} - {G_{1{\phi}} G_{{\phi}2} \over G_{{\phi}1} + G_{{\phi}2}}.
\end{eqnarray}

Using unitarity of $S$ we may eliminate the conductance coefficients $G_{kl}$ which involve the voltage probe,
\begin{eqnarray} 
  G &=& - G_{12} + {(G_{11} + G_{12})(G_{22} + G_{12}) \over G_{11} + G_{12} + G_{21} + G_{22}}.
  \label{eq:conductance}
\end{eqnarray}
The remaining conductance coefficients are constructed from the matrix
\begin{equation} \label{eq:Ssub}
  \tilde S = \left( \begin{array}{ccc} s_{11} & s_{12} \\
                      s_{21} & s_{22} \end{array} \right),
\end{equation}
which formally represents the scattering matrix of an absorbing system. The first term in Eq.\ (\ref{eq:conductance}) would be the conductance if the voltage probe would truly absorb the electrons which enter it. The second term accounts for the electrons that are reinjected from the phase-breaking reservoir, thereby ensuring particle conservation in the voltage-probe model.

The imaginary potential model relates $\tilde S$ to a Hamiltonian $\tilde H$ with a spatially uniform, negative imaginary potential $-i \gamma \Delta/4 \pi$. As used in Refs.\ \onlinecite{Efetov1995} and \onlinecite{McCannLerner}, it retains only the first term in Eq.\ (\ref{eq:conductance}), and therefore does not conserve particles. We correct this by including the second term. We will now show that this particle-conserving imaginary potential model is equivalent to the voltage-probe model in the limit $N_{\phi} \to \infty$, $\Gamma_{\phi} \to 0$, $N_{\phi} \Gamma_{\phi} \equiv \gamma$.

Our equivalence proof is based on the general relationship\cite{VWZ,IWZ}
\begin{equation}
  \tilde S = 1 - 2 \pi i \tilde W^{\dagger} (E_F - \tilde H + i \pi \tilde W \tilde W^{\dagger})^{-1} \tilde W \label{eq:SWH}
\end{equation}
between the $N \times N$ scattering matrix $\tilde S$ ($N=N_1+N_2$) and the $N' \times N'$ Hamiltonian $\tilde H$ (the limit $N' \to \infty$ is taken later on). The Hamiltonian contains an imaginary potential, $\tilde H_{\mu \nu} = H_{\mu \nu} - i \delta_{\mu \nu} \gamma \Delta/4 \pi$, with $H$ a Hermitian matrix. For a chaotic cavity, $H$ is taken from the Gaussian ensemble of random matrix theory.\cite{Mehta} The $N' \times N$ matrix $\tilde W$ has elements\cite{IWZ,Brouwer1995}
\begin{equation}
  \pi \tilde W_{\mu n}^2 = \pi^{-1}\delta_{\mu n} N' \Delta \left(2 \Gamma_n^{-1} - 1 - 2 \Gamma_n^{-1} \sqrt{1 - \Gamma_n} \right). \label{eq:W}
\end{equation}
Here $\Gamma_n$ is the transmission probability of mode $n$ in the leads and the energy $\Delta$ is the mean level spacing of $H$. We embed $\tilde W$ into an $N' \times N'$ matrix by the definition $\tilde W_{\mu n} = 0$ for $N < n \le N'$, and define
\begin{equation}
 \pi W_{\mu n}^2 = \pi \tilde W_{\mu n}^2 + \delta_{\mu n} \gamma \Delta/4 \pi. \label{eq:Wsub}
\end{equation}
Substitution into Eq.\ (\ref{eq:SWH}) shows that $\tilde S$ is an $N \times N$ submatrix of an $N' \times N'$ unitary matrix
\begin{equation}
  S = 1 - 2 \pi i W^{\dagger} (E_F - H + i \pi W W^{\dagger})^{-1} W. \label{eq:SWH2}
\end{equation}
We have neglected the difference between $\tilde W_{\mu \mu}$ and $W_{\mu \mu}$ for $1 \le \mu \le N$, which is allowed in the limit $N' \to \infty$. The matrix $S$ is the scattering matrix of a cavity with three leads: Two real leads with $N_1$, $N_2$ modes, plus a fictitious lead with $N'-N$ modes. The transmission probability $\Gamma_{n}$ of a mode in the fictitious lead follows from Eqs.\ (\ref{eq:W}) and (\ref{eq:Wsub}),
\begin{equation}
  \Gamma_{n} = {4 \pi^2 W_{nn}^2 N' \Delta \over (N' \Delta + \pi^2 W_{nn}^2)^2} \to {\gamma \over N'} \ \mbox{if $N' \to \infty$},
\end{equation}
where we have used that $\pi W_{nn}^2 = \gamma \Delta/4 \pi$ for $N < n \le N'$.

We conclude that the particle-conserving imaginary potential model and the voltage-probe model are equivalent in the limit $N_{\phi} = N' - N\to \infty$, $\Gamma_{\phi} = \gamma/N'\to 0$, $N_{\phi} \Gamma_{\phi} = \gamma (1 - N/N') \to \gamma$.

\section{Single-mode point contacts} \label{sec:3}

The effect of quantum-interference on the conductance is maximal if the point contacts which couple the chaotic cavity to the source and drain reservoirs have only a single propagating mode at the Fermi level. Then the sample-to-sample fluctuations of the conductance are of the same size as the average conductance itself. One thus needs the entire conductance distribution to characterize an ensemble of quantum dots. (An ensemble may be generated by small variations in shape or in Fermi energy.) 

In the absence of dephasing, the conductance distribution $P(G)$ is strongly non-Gaussian.\cite{PrigodinEfetovIida,BarangerMello,JPB,BrouwerBeenakker1994} For ideal point contacts (transmission probabilities $\Gamma_1 = \Gamma_2 = 1$), one finds\cite{BarangerMello,JPB}
\begin{equation}
  P(G) = \case{1}{2} \beta G^{(\beta-2)/2}.
\end{equation}
The symmetry parameter $\beta=2$ ($1$) in the presence (absence) of a time-reversal-symmetry breaking magnetic field. For high tunnel barriers ($\Gamma_1, \Gamma_2 \ll 1$), $P(G)$ is maximal for $G = 0$, and drops off $\propto G^{-3/2}$ for $G \gg \Gamma_1 \Gamma_2$.\cite{PrigodinEfetovIida,BrouwerBeenakker1994} In this section, we compute the conductance distribution in the presence of dephasing, using the voltage-probe model in the limit $N_{\phi} \to \infty$, $\Gamma_{\phi} \to 0$ at fixed $N_{\phi} \Gamma_{\phi}$, in which it is equivalent to the current-conserving imaginary potential model. We focus on the case of ideal point contacts, and discuss the effect of tunnel barriers briefly at the end of the section.

The scattering matrix $S$ is distributed according to the Poisson kernel,\cite{Brouwer1995,Hua,MPS,BarangerMello1996}
\begin{equation} \label{eq:poisson1}
  P(S) = {1 \over V} {\det(1-\bar S \bar S^{\dagger})^{(\beta M + 2-\beta)/2} \over
                      |\det(1-\bar S S^{\dagger})|^{\beta M + 2 - \beta}},
\end{equation}
where $V$ is a normalization constant, $M=N_1+N_2+N_{\phi}$ is the dimension of $S$, and $\bar S$ is a diagonal matrix with diagonal elements $\bar S_{nn} = \sqrt{1 - \Gamma_{n}}$. Here $\Gamma_n$ is the transmission probability of mode $n$ ($\Gamma_n \equiv \Gamma_{\phi}$ for $N_1 + N_2 < n \le M$). The measure $dS$ is the invariant measure on the manifold of unitary (unitary symmetric) matrices for $\beta=2$ ($1$). 

We now specialize to the case of ideal single-mode point contacts, $N_1 = N_2 = 1$ and $\Gamma_1 = \Gamma_2 = 1$. We seek the distribution of the $2 \times 2$ submatrix $\tilde S$ defined in Eq.\ (\ref{eq:Ssub}). We start with the polar decomposition of $S$,
\begin{equation} \label{eq:Spolar}
  S = \left( \begin{array}{ll} u  & 0 \\ 0 & v  \end{array} \right)
      \left( \begin{array}{cc} \sqrt{1-t^{\dagger} t} & i t^{\dagger}\\ 
                               i t & \sqrt{1-tt^{\dagger}} \end{array} \right)
      \left( \begin{array}{ll} u' & 0 \\ 0 & v' \end{array} \right),
\end{equation}
where $u$ and $u'$ ($v$ and $v'$) are $2 \times 2$ ($N_{\phi} \times N_{\phi}$) unitary matrices, and $t$ is a $N_{\phi} \times 2$ matrix with all elements equal to zero except $t_{nn} = \sqrt{T_n}$, $n=1,2$. In the presence of time-reversal symmetry, $u' = u^{\rm T}$ and $v' = v^{\rm T}$. In terms of the polar decomposition (\ref{eq:Spolar}) we have 
\begin{equation}
  \tilde S = u \left( \begin{array}{cc} \sqrt{1-T_1} & 0 \\ 0 & \sqrt{1 - T_2} \end{array} \right) u'.
\end{equation} 
The two parameters $T_1$ and $T_2$ govern the strength of the absorption by the voltage probe. For $T_1, T_2 \to 0$ the matrix $\tilde S$ is unitary and there is no absorption, whereas for $T_1, T_2 \to 1$ the matrix $\tilde S$ vanishes and the absorption is complete. Substitution of the invariant measure\cite{BeenakkerReview}
\begin{eqnarray} 
  dS &=& |T_1 - T_2|^{\beta} (T_1 T_2)^{(\beta N_{\phi} - 2 - \beta)/2}\nonumber \\ && \mbox{} \times   du du' dv dv' dT_1 dT_2 \label{eq:mupolar}
\end{eqnarray}
and the polar decomposition (\ref{eq:Spolar}) into the Poisson kernel (\ref{eq:poisson1}), yields the distribution of $\tilde S$ in the form
\begin{mathletters} \label{eq:PS}
\begin{eqnarray}
  && P(T_1,T_2,u,u') =
   \Gamma_{\phi}^{N_{\phi}(\beta N_{\phi}+2+\beta)/2} |T_1 - T_2|^{\beta}
   \nonumber \\ && \ \ \mbox{} \times
   {1 \over V} \int dv \int dv'\, {(T_1 T_2)^{(\beta N_{\phi}-2 - \beta)/2} \over
   |\det(1- v'v\, \tau)|^{\beta N_{\phi} + 2 + \beta}},  
  ~~~ \label{eq:PSa} \\
 && \tau = \sqrt{(1-\Gamma_{\phi})(1-tt^{\dagger})}. \label{eq:PSb}
\end{eqnarray}
\end{mathletters}%
Since Eq.\ (\ref{eq:PS}) is independent of $u$ and $u'$, the matrices $u$ and $u'$ are uniformly distributed in the unitary group, and the distribution of $\tilde S$ is completely determined by the joint distribution $P(T_1,T_2)$ of the absorption probabilities $T_1$ and $T_2$. 

It remains to perform the integral over $v$ and $v'$ in Eq.\ (\ref{eq:PS}). This is a non-trivial calculation, which we describe in the appendix. The final result in the limit $N_{\phi} \to \infty$, $\Gamma_{\phi} \to 0$ at fixed $\gamma = N_{\phi} \Gamma_{\phi}$ is
\bleq
\begin{mathletters} \label{eq:PT}
\begin{eqnarray}
  P(T_1,T_2) &=&
  \case{1}{8} T_1^{-4} T_2^{-4}
  \exp\left[-\case{1}{2}\gamma(T_1^{-1} + T_2^{-1}) \right] |T_1-T_2| 
  \left[ \gamma^2(2 - 2 e^{\gamma} + \gamma + \gamma e^{\gamma}) 
\right. \nonumber \\ && \left. \mbox{}
-
     \gamma (T_1 + T_2) (6 - 6 e^{\gamma} + 4 \gamma + 2 \gamma e^{\gamma} + \gamma^2) 
+
     T_1 T_2 (24 - 24 e^{\gamma} + 18 \gamma + 6 \gamma e^{\gamma} + 6 \gamma^2 + \gamma^3)
  \right]  \label{eq:Pb1}
\end{eqnarray}
for $\beta=1$ (presence of time-reversal symmetry), and
\begin{eqnarray}
  P(T_1,T_2) &=& 
  \case{1}{2} T_1^{-6} T_2^{-6} \exp\left[-\gamma(T_1^{-1} + T_2^{-1})\right] (T_1 - T_2)^{2}
  \nonumber \\ && \mbox{} \times \left[
  \gamma^4(1 - 2 e^{\gamma} + e^{2 \gamma} - \gamma^2 e^{\gamma})
  - \gamma^3 (T_1 + T_2) (4 - 8 e^{\gamma} + 4 e^{2 \gamma} + 2 \gamma 
                            - 2 \gamma e^{\gamma} -2 \gamma^2 e^{\gamma} - \gamma^3 e^{\gamma})
  \nonumber \right. \\ && \left. \mbox{}
  + \gamma^2 (T_1^2 + T_2^2) (2 - 4 e^{\gamma} + 2 e^{2 \gamma} + 4 \gamma 
                                - 4 \gamma e^{\gamma} + \gamma^2 + \gamma^2 e^{\gamma} 
                                - \gamma^3 e^{\gamma})
  \nonumber \right. \\ && \left. \mbox{}
  + \gamma^2 T_1 T_2 (20 - 40 e^{\gamma} + 20 e^{2 \gamma} + 16 \gamma 
                       - 16 \gamma e^{\gamma} + 4 \gamma^2 - 8 \gamma^2 e^{\gamma}
                       - 4 \gamma^3 e^{\gamma} - \gamma^4 e^{\gamma})
  \nonumber \right. \\ && \left. \mbox{}
  - \gamma T_1 T_2 (T_1 + T_2) (12 - 24 e^{\gamma} + 12 e^{2 \gamma} + 24 \gamma 
                                   - 24 \gamma e^{\gamma} + 12 \gamma^2 + 2 \gamma^3
                                   - 2 \gamma^3 e^{\gamma}- \gamma^4 e^{\gamma})
  \nonumber \right. \\ && \left. \mbox{}
  + T_1^2 T_2^2 (12 - 24 e^{\gamma} + 12 e^{2 \gamma} + 24 \gamma - 24 \gamma e^{\gamma} 
                    + 24 \gamma^2 - 12 \gamma^2 e^{\gamma} + 8 \gamma^3
                    + 4 \gamma^3 e^{\gamma} + \gamma^4 - 2 \gamma^4 e^{\gamma})
  \right] \label{eq:Pb2}
\end{eqnarray}
\end{mathletters}%
\eleq
for $\beta=2$ (absence of time-reversal symmetry).

\begin{figure}

\epsfxsize=0.95\hsize
\epsffile{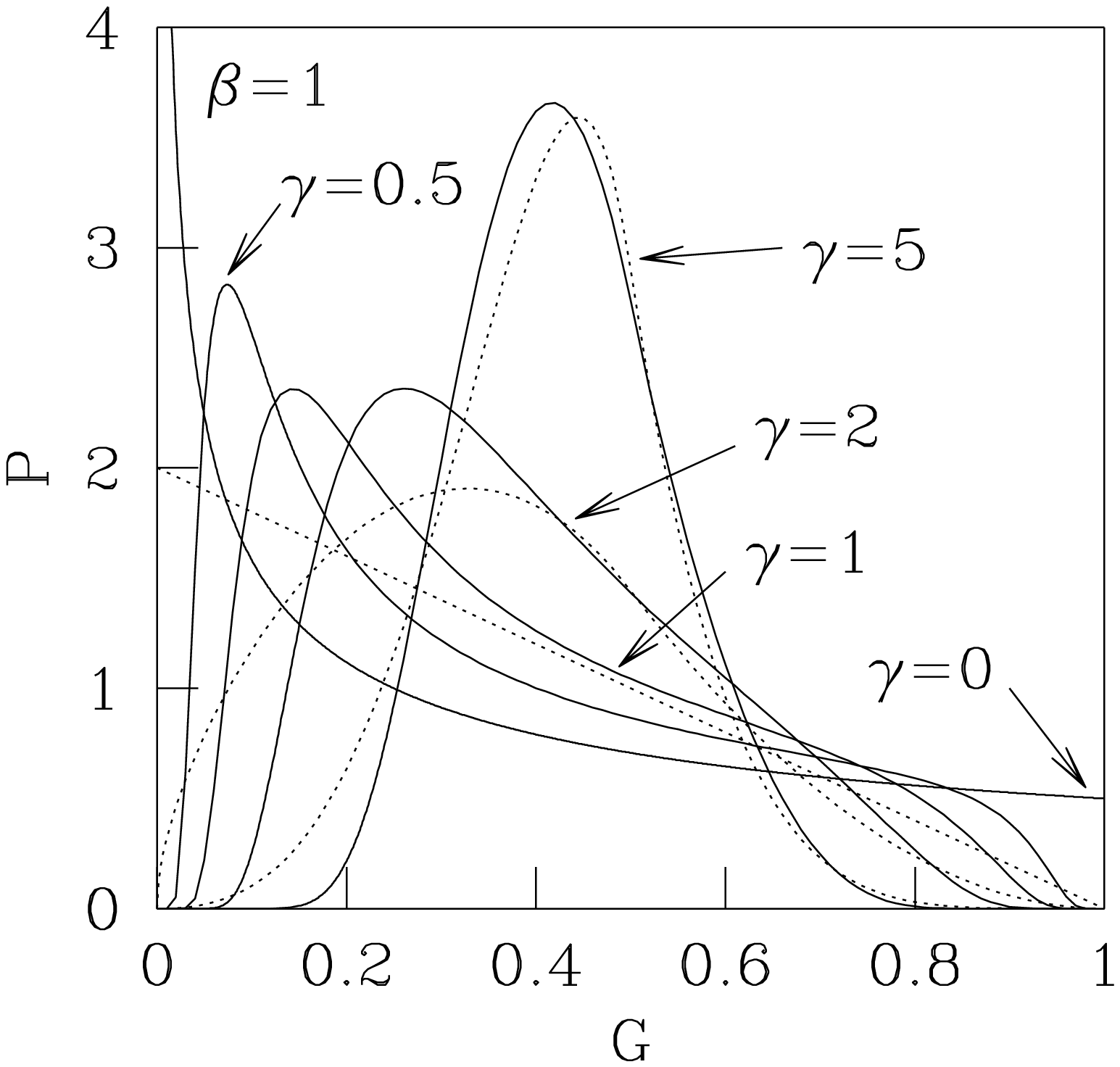}
\vspace{-1.9cm}

\epsfxsize=0.95\hsize
\epsffile{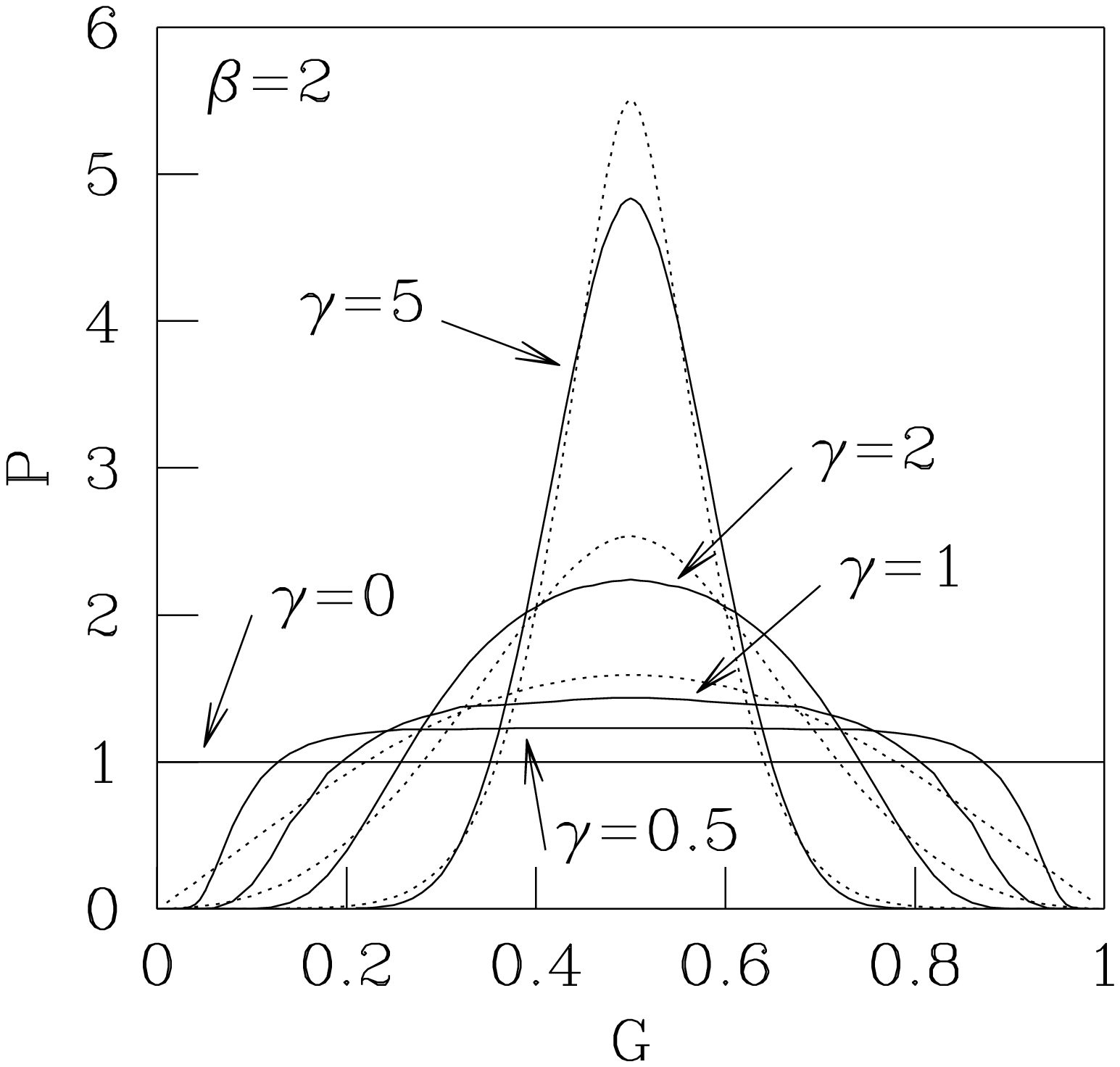}
\vspace{0.3cm}
\caption{\label{fig:2}
Solid curves: Conductance distributions of a quantum dot with two ideal single-mode point contacts, computed from Eqs.\ (\protect\ref{eq:PT}) and (\protect\ref{eq:Gueq}) for dephasing rates $\gamma=0$, $0.5$, $1$, $2$, and $5$. The top panel is for zero magnetic field ($\beta=1$), the bottom panel for broken time-reversal symmetry ($\beta=2$). The dotted curves are the results of Refs.\ \protect\onlinecite{BarangerMello1995} and \protect\onlinecite{BrouwerBeenakker1995} for the model of an ideal voltage probe (without a tunnel barrier), in which dephasing is not fully uniform in phase space. For $\gamma=0$ the two models coincide. The value $\gamma=0.5$ is not accessible in the model of an ideal voltage probe (because $\gamma = N_{\phi} \Gamma_{\phi}$ can take on only integer values if $\Gamma_{\phi} = 1$).}
\end{figure}

\begin{figure}

\epsfxsize=0.95\hsize
\epsffile{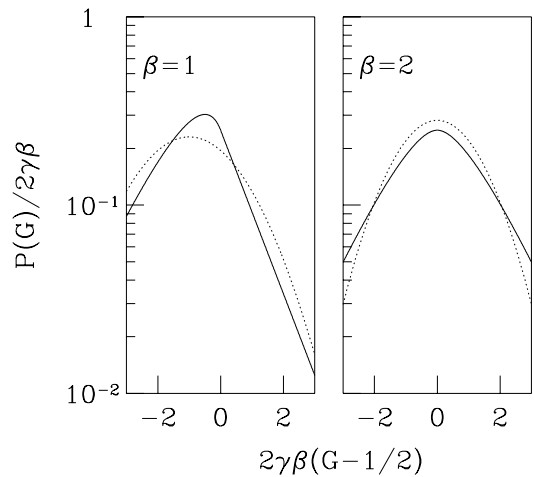}
\vspace{0.3cm}

\caption{\label{fig:2d} The limiting conductance distribution (\protect\ref{eq:PGdistr}) for $\gamma \gg 1$ (solid curves). A Gaussian distribution with the same mean and variance is shown for comparison (dotted curves).}
\end{figure}

To relate the conductance $G$ to $T_1$, $T_2$, $u$, and $u'$, we substitute the polar decomposition of $S$ into Eq.\ (\ref{eq:conductance}), with the result
\begin{eqnarray}
  G &=& \sum_{i,j=1}^{2} u^{\vphantom{2}}_{1i} u'_{i2} u_{1j}^{*} u_{j2}'^{*} \sqrt{(1-T_i)(1-T_j)}
      \nonumber \\ && \mbox{} +
      \left( T_1 + T_2 \right)^{-1} \sum_{i,j=1}^{2} |u_{1i}^{\vphantom{2}}|^2 |u'_{j2}|^2 T_i T_j. \label{eq:Gueq}
\end{eqnarray} 
Eqs.\ (\ref{eq:PT}) and (\ref{eq:Gueq}), together with the uniform distribution of the $2 \times 2$ matrices $u$, $u'$ over the unitary group, fully determine the distribution $P(G)$ of the conductance of a chaotic cavity with two ideal single-mode point contacts. We parameterize $u$, $u'$ in Euler angles and obtain $P(G)$ as a four-dimensional integral, which we evaluate numerically. The distribution is plotted in Fig.\ \ref{fig:2} (solid curves) for several values of the dimensionless dephasing rate $\gamma = 2 \pi \hbar/\tau_{\phi} \Delta$. For $\gamma \gg 1$,\cite{footnote} the conductance distribution becomes peaked around the classical conductance $G = 1/2$,
\begin{eqnarray}
  P(G) \to {\gamma \beta \over 2} \left( 1 + |x| - \delta_{\beta 1} x \right) e^{-|x|}\ \ \mbox{if $\gamma \gg 1$}, \label{eq:PGdistr}
\end{eqnarray}
where $x = 2 \gamma \beta (G - 1/2)$. Notice that the distribution remains non-Gaussian for all values of $\gamma$. The limiting distribution (\ref{eq:PGdistr}) is plotted in Fig.\ \ref{fig:2d}, for $\beta=1$ and $2$. The average and variance of the conductance are
\begin{mathletters} \label{eq:PGavgvar}
\begin{eqnarray}
  \langle G \rangle &=& \case{1}{2} - \case{1}{2} \delta_{\beta1} \gamma^{-1} +
     {\cal O}(\gamma^{-2}), \\
  \mbox{var}\, G &=& \case{1}{4}(1 + 2 \delta_{\beta1}) \gamma^{-2} + {\cal O}(\gamma^{-3}).
\end{eqnarray}
\end{mathletters}%

The effect of dephasing was previously studied in Refs.\ \onlinecite{BarangerMello1995} and \onlinecite{BrouwerBeenakker1995} for the case $\Gamma_{\phi} = 1$ of an ideal voltage probe (without a tunnel barrier). The corresponding results are also shown in Fig.\ \ref{fig:2} (dotted curves). We see that the limit $N_{\phi} \to \infty$, $\Gamma_{\phi} \to 0$ results in narrower distributions at the same value of $\gamma=N_{\phi} \Gamma_{\phi}$. In particular, the tails $G \to 0$ and $G \to 1$ are strongly suppressed even for the smallest $\gamma$, in contrast with the case of the ideal voltage probe. The physical reason for the difference is that keeping $N_{\phi}$ small and setting $\Gamma_{\phi}$ equal to $1$ corresponds to dephasing which is not fully uniform in phase space, and therefore not as effective as the limit $N_{\phi} \to \infty$, $\Gamma_{\phi} \to 0$. For large $\gamma$, the difference vanishes, and the distribution (\ref{eq:PGdistr}) is recovered for an ideal voltage probe as well. (The fact that the conductance fluctuations around $G=1/2$ are non-Gaussian was overlooked in Refs.\ \onlinecite{BarangerMello1995} and \onlinecite{BrouwerBeenakker1995}.)

\begin{figure}
\vspace{-1cm}

\epsfxsize=0.95\hsize
\epsffile{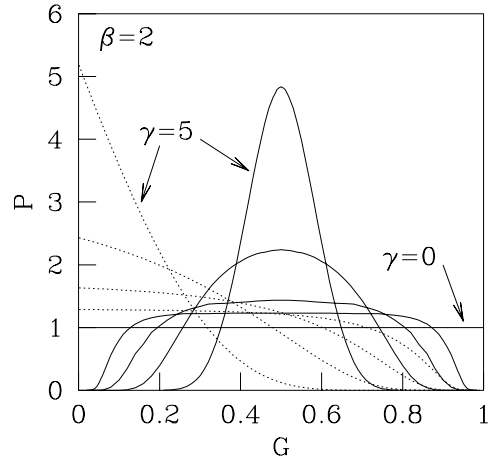}
\caption{\label{fig:2c} Solid curves: Same as in Fig.\ \protect\ref{fig:2}, bottom panel. Dotted curves: Results of the imaginary potential model without particle conservation.}
\end{figure}

We have shown in the previous section that the voltage-probe model in the limit $N_{\phi} \to \infty$, $\Gamma_{\phi} \to 0$ is equivalent to the particle-conserving imaginary potential model. The requirement of particle conservation is essential. This is illustrated in Fig.\ \ref{fig:2c}, where we compare our results with those obtained from the imaginary potential model without enfor\-cing conservation of particles. [This model corresponds to setting $G = -G_{12}$ in Eq.\ (\ref{eq:conductance}) and was first solved in Ref.\ \onlinecite{PrigodinEfetovIida}.] For $\gamma \gg 1$, the imaginary potential without particle conservation yields a distribution which is maximal at $G = 0$, instead of a strongly peaked distribution around $G=1/2$ [cf.\ Eq.\ (\ref{eq:PGdistr})].

\begin{figure}

\epsfxsize=0.95\hsize
\epsffile{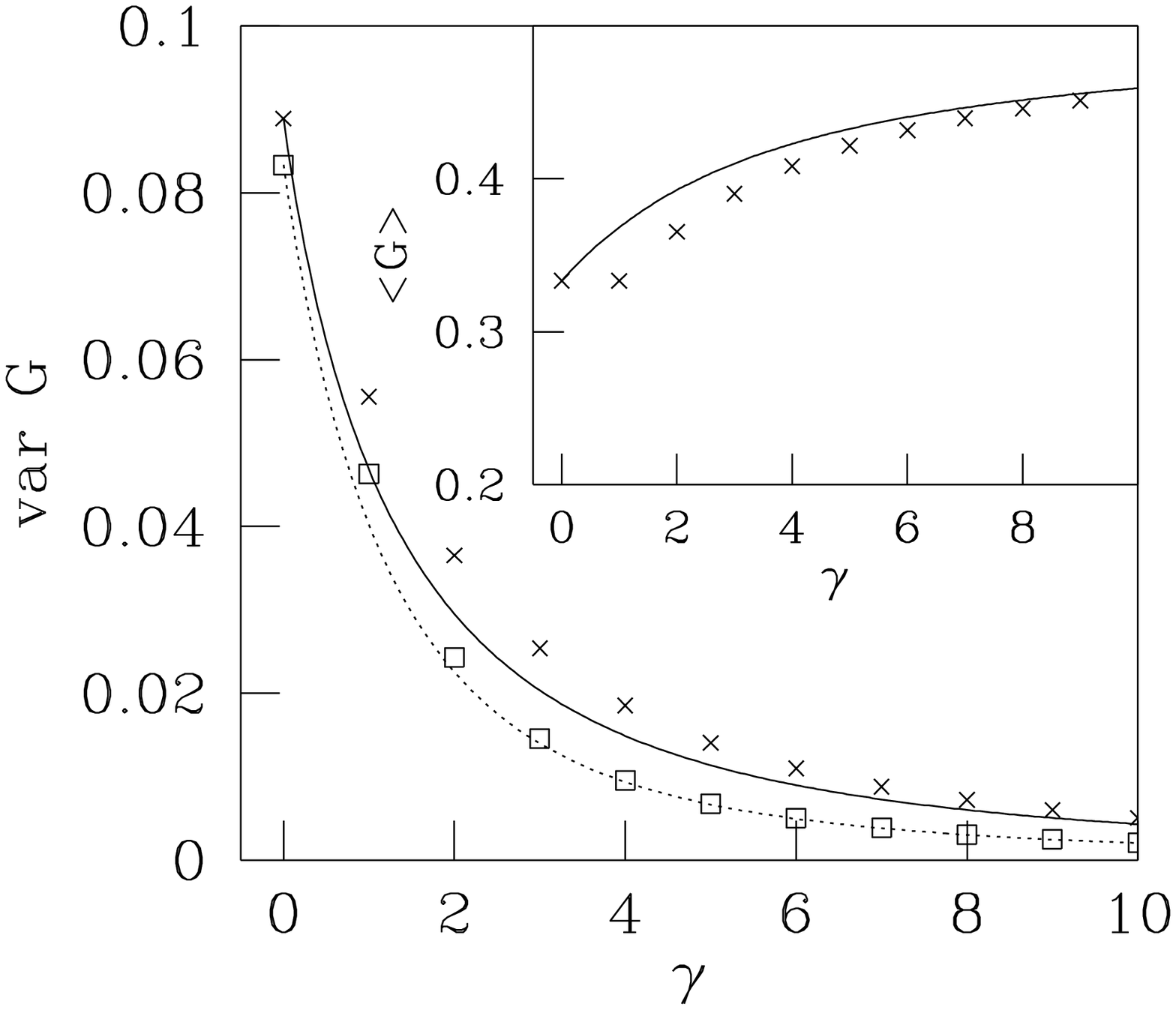}

\caption{\label{fig:3} Variance of the conductance as a function of the dephasing rate $\gamma$, for $\beta=1$ (solid curve) and $\beta=2$ (dotted curve), computed from Eqs.\ (\protect\ref{eq:PT}) and (\protect\ref{eq:Gueq}). The crosses ($\beta=1$) and squares ($\beta=2$) at integer $\gamma$ result from the model of Refs.\ \protect\onlinecite{BarangerMello1995} and \protect\onlinecite{BrouwerBeenakker1995} with the ideal voltage probe. The inset shows the average conductance for $\beta=1$. (For $\beta=2$ the average is trivially equal to $1/2$ for all $\gamma$ in both models.)}
\end{figure}

The first two moments of the conductance can be computed analytically from Eqs.\ (\ref{eq:PT}) and (\ref{eq:Gueq}). The resulting expressions (which are too lengthy to report here) are plotted in Fig.\ \ref{fig:3}. The markers at integer values of $\gamma$ are the results of the ideal voltage-probe model of Refs.\ \onlinecite{BarangerMello1995} and \onlinecite{BrouwerBeenakker1995}, where $\Gamma_{\phi} = 1$ and $\gamma=N_{\phi} = 0,1,2,\ldots$. The remarkable result\cite{BrouwerBeenakker1995} that $\langle G \rangle$ is the same for $\gamma=0$ and $\gamma=1$ is special for dephasing by a single-mode voltage probe: The present model with spatially uniform dephasing has a strictly monotonic increase of $\langle G \rangle$ with $\gamma$ for $\beta=1$.

Sofar we have considered ideal point contacts. Non-ideal point contacts (i.e.\ point contacts with tunnel barriers) correspond to $\Gamma_1, \Gamma_2 < 1$ in the distribution (\ref{eq:poisson1}) of $S$. This case can be mapped onto that of ideal point contacts by the parameterization\cite{Brouwer1995,Hua,MPS}
\begin{eqnarray}
  S &=& R + T (1 - S' R)^{-1} S' T, \label{eq:Smap}
\end{eqnarray}
where $R$ and $T = i \sqrt{1 - R^2}$ are diagonal matrices. The only nonzero elements of $R$ are $R_{11} = \sqrt{1 - \Gamma_1}$ and $R_{22} = \sqrt{1 - \Gamma_2}$. The distribution of $S'$ is given by the Poisson kernel (\ref{eq:poisson1}) with $\Gamma_{1} = \Gamma_{2} = 1$. Physically, $S'$ is the scattering matrix of the quantum dot without the tunnel barriers in the point contacts, while $R$ ($T$) is the reflection (transmission) matrix of the tunnel barriers in the absence of the quantum dot.\cite{Brouwer1995} We may restrict the parameterization (\ref{eq:Smap}) to the $2 \times 2$ submatrix $\tilde S$,
\begin{eqnarray} \label{eq:StildeRT}
  \tilde S &=& \tilde R + \tilde T (1 - \tilde S' \tilde R)^{-1} \tilde S' \tilde T,
\end{eqnarray}
where the matrices $\tilde S'$, $\tilde R$, and $\tilde T$ are the upper-left $2 \times 2$ submatrices of $S'$, $R$, and $T$, respectively. The matrix $\tilde S'$ has the distribution given by Eqs.\ (\ref{eq:PS}) and (\ref{eq:PT}). The matrices $\tilde R$ and $\tilde T$ are fixed, so the distribution of $\tilde S$ follows directly from Eq.\ (\ref{eq:StildeRT}).

For strong dephasing ($\gamma \gg \Gamma_1, \Gamma_2$) we find that the conductance distribution becomes a Gaussian with mean and variance given by
\begin{mathletters} \label{eq:tunnelGauss}
\begin{eqnarray}
  \langle G \rangle &=& {\Gamma_1 \Gamma_2 \over \Gamma_1 + \Gamma_2} - {2 \Gamma_1^2 \Gamma_2^2 (4/\beta - \Gamma_1 - \Gamma_2) \over \gamma (\Gamma_1 + \Gamma_2)^3}, \\
  \mbox{var}\, G &=& {4 \Gamma_1^2 \Gamma_2^2 (\Gamma_1^2 + \Gamma_2^2 - \Gamma_1 \Gamma_2^2 - \Gamma_1^2 \Gamma_2) \over \beta \gamma (\Gamma_1 + \Gamma_2)^3}. \label{eq:tunnelGaussb}
\end{eqnarray}
\end{mathletters}%
The average conductance $\langle G \rangle$ is the classical series conductance of the two point-contact conductances $\Gamma_1$ and $\Gamma_2$. Fluctuations around the classical conductance are of order $\gamma^{-1/2}$. For ideal point contacts ($\Gamma_{1}, \Gamma_{2} \to 1$) the variance (\ref{eq:tunnelGaussb}) vanishes. The higher-order fluctuations are non-Gaussian, described by Eq.\ (\ref{eq:PGdistr}). 

Again our result is entirely different from that of the imaginary potential model without particle conservation,\cite{PrigodinEfetovIida,McCannLerner} where $P(G)$ becomes sharply peaked at $G=0$ when $\gamma \gg \Gamma_1,\Gamma_2$. We have verified that we recover the results of Ref.\ \onlinecite{PrigodinEfetovIida} from our Eqs.\ (\ref{eq:PT}) and (\ref{eq:Gueq}) if we retain only the first term in Eq.\ (\ref{eq:conductance}), i.e.\ if we set $G=-G_{12}$. The results of Ref.\ \onlinecite{McCannLerner} are recovered if we symmetrize this term, i.e.\ if we set $G = -(G_{12} + G_{21})/2$. (This is different from $-G_{12}$ if $\beta=2$ and $\gamma \neq 0$.) Once particle conservation is enfor\-ced, the imaginary potential model leads unambiguously to Eq.\ (\ref{eq:tunnelGauss}).

\section{Multi-mode point contacts} \label{sec:4}

In this section we consider the case $N_1, N_2 \gg 1$ of a large number of modes in the two point contacts. The conductance distribution is then a Gaussian, hence it suffices to compute the first two moments of $G$. We first consider ideal point contacts ($\Gamma_1 = \Gamma_2 = 1$), and discuss the effect of tunnel barriers at the end. 

For $N_1, N_2 \gg 1$ the integration over the scattering matrix $S$ with the probability distribution (\ref{eq:poisson1}) can be done using the diagrammatic technique of Ref.\ \onlinecite{BrouwerBeenakker1996}. The result for the average of the conductance coefficients $G_{ij}$ is
\begin{mathletters} \label{eq:Ravg}
\begin{eqnarray}
&&  \langle G_{ij} \rangle = N_i \delta_{ij} - {N_i N_j \over N + N_{\phi} \Gamma_{\phi}}
   + \delta_{\beta,1} A_{ij}, \\
&&  A_{ij} = {N_i N_j (N + 2 N_{\phi} \Gamma_{\phi} - N_{\phi}\Gamma_{\phi}^2) \over (N+N_{\phi} \Gamma_{\phi})^3} - {\delta_{ij}N_i \over N+N_{\phi} \Gamma_{\phi}},~~~~~
\end{eqnarray}
\end{mathletters}%
up to terms of order $N^{-1}$. (We recall that $N=N_1+N_2$.) For the covariances $\mbox{cov}\,(G_{ij},G_{kl}) \equiv \langle G_{ij} G_{kl} \rangle - \langle G_{ij} \rangle \langle G_{kl} \rangle$ we find
\begin{eqnarray}
 &&  \mbox{cov}\,(G_{ij},G_{kl}) = A_{ik} A_{jl} + \delta_{\beta,1} A_{il} A_{jk} \ \ \ \ \ \nonumber \\ && \ \ \ \ \ \ \ \ \ \ \mbox{} + {2N_i N_j N_k N_l N_{\phi} (N_{\phi}+N) \Gamma_{\phi}^2 (1-\Gamma_{\phi}) \over \beta(N+N_{\phi} \Gamma_{\phi})^6}.~~~~ \label{eq:Rcov}
\end{eqnarray}
In order to find the average and variance of the conductance in the presence of dephasing, we substitute Eqs.\ (\ref{eq:Ravg}) and (\ref{eq:Rcov}) into Eq.\ (\ref{eq:conductance}). The result is
\begin{mathletters} \label{eq:GINEL}
\begin{eqnarray}
  \langle G \rangle &=& {N_1 N_2 \over N} \left( 1 -{\delta_{\beta1} \over N+\gamma} \right), \label{eq:GavgINEL} \\
  \mbox{var}\, G &=& {2 N_1^2 N_2^2 \over \beta N^2 (N+\gamma)^2}, \label{eq:GvarINEL} 
\end{eqnarray}
\end{mathletters}%
with $\gamma = N_{\phi} \Gamma_{\phi}$. 

Eq.\ (\ref{eq:GavgINEL}) was previously obtained by Aleiner and Larkin.\cite{AL} Eq.\ (\ref{eq:GvarINEL}) for $\mbox{var}\, G$ agrees with the interpolation formula of Baranger and Mello\cite{BarangerMello1995}. The present derivation shows that this interpolation formula is in fact a rigorous result of perturbation theory. [However, the interpolation formula of Ref.\ \onlinecite{BarangerMello1995} for $\langle G \rangle$ differs from Eq.\ (\ref{eq:GavgINEL}).] In the final expression for $\langle G \rangle$ and $\mbox{var}\, G$ only the product $N_{\phi} \Gamma_{\phi}$ appears, although the moments of the conductance coefficients $G_{ij}$ depend on $N_{\phi}$ and $\Gamma_{\phi}$ separately. Apparently, in large-$N$ perturbation theory the precise choice of $N_{\phi}$ and $\Gamma_{\phi}$ in the voltage-probe model is irrelevant, the conductance distribution being determined by the product $N_{\phi} \Gamma_{\phi}$ only.
For small dephasing rates $\gamma \ll N$, Eq.\ (\ref{eq:GINEL}) agrees with Efetov's result,\cite{Efetov1995} who used the imaginary potential model without enfor\-cing particle conservation. However, for $\gamma \gtrsim N$, our result differs from that of Ref.\ \onlinecite{Efetov1995}, indicating the importance of particle conservation once the dephasing rate $\gamma$ and the dimensionless escape rate $N$ through the point contacts become comparable.

We have carried out the same calculation for the case of non-ideal point contacts. The transmission probability of mode $n$ is denoted by $\Gamma_n$ ($n=1,\ldots,N_1$ corresponding to the first point contact, $n=N_1+1,\ldots,N_1+N_2$ to the second point contact). The result is
\begin{mathletters} \label{eq:GtunnelMom}
\begin{eqnarray}
  \langle G \rangle &=& {g_1^{\vphantom{2}} g_1' \over g} - \delta_{\beta 1}
    {g_2^{\vphantom{2}} g_1'^2 + g_1^2 g_2' \over g^2 (g + \gamma)}, \\
  \mbox{var}\, G &=& {4 g_1^{2} g_1'^{2} \over \beta g^{2} (g + \gamma)^2}
   + {4 (g_1^{4} g_2' - g_1^{4} g_3' + g_2 g_1'^{4} - g_3 g_1'^{4}) \over \beta g^4 (g + \gamma)}
   \nonumber \\ && \mbox{}
   + {6 (g_1^{2} g_2'^{} + g_2^{} g_1'^{2})^2 \over \beta g^4 (g + \gamma)^2}
   - {8 g_1 g_1' (g_2 g_1'^{2} + g_2'g_1^{2}) \over \beta g^3 (g+\gamma)^2}
   , \nonumber \\ ~ \\
  g_p^{\vphantom{2}} &=& \sum_{n=1}^{N_1} \Gamma_n^p, \ \ 
  g_p' = \sum_{n=1+N_1}^{N_1+N_2} \Gamma_n^p, \ \
  g = g_1 + g_1'.
\end{eqnarray}
\end{mathletters}%
One can check that Eq.\ (\ref{eq:GtunnelMom}) reduces to Eq.\ (\ref{eq:GINEL}) for ideal point contacts (when $g_{p} = N_1$, $g_{p}' = N_2$). As in the case of single-mode point contacts, $\mbox{var}\, G \propto \gamma^{-2}$ for $\gamma \gg 1$ without tunnel barriers, while $\mbox{var}\, G \propto \gamma^{-1}$ otherwise.

\section{Conclusion} \label{sec:5}

In summary, we have demonstrated the equivalence of two models for dephasing, the voltage-probe model and the imaginary potential model. In doing so we have corrected a number of shortcomings of each model, notably the non-uniformity of the dephasing in the voltage-probe model of Refs.\ \onlinecite{BarangerMello1995} and \onlinecite{BrouwerBeenakker1995} and the lack of particle conservation in the imaginary potential model of Refs.\ \onlinecite{Efetov1995} and \onlinecite{McCannLerner}. We have calculated the distribution of the conductance and shown that it peaks at the classical conductance for strong dephasing once particle conservation is enfor\-ced, thereby reconciling the contradictory results of Refs.\  \onlinecite{BarangerMello1995} and \onlinecite{BrouwerBeenakker1995}, on the one hand, and Refs.\ \onlinecite{Efetov1995} and \onlinecite{McCannLerner}, on the other hand. 
We find that for ideal single-mode point contacts (no tunnel barriers), conductance fluctuations are non-Gaussian and $\propto \tau_{\phi}$ for strong dephasing ($\tau_{\phi} \to 0$). In the case of non-ideal point contacts (with tunnel barriers), fluctuations are larger ($\propto \sqrt{\tau_{\phi}}$) and Gaussian for $\tau_{\phi} \to 0$ . 

The effect of dephasing becomes appreciable when the dimensionless dephasing rate $\gamma = 2 \pi \hbar/\tau_{\phi} \Delta$ is of the same order as the dimensionless escape rate $g = \sum_{n} \Gamma_{n}$ through the two point contacts. For $\gamma \gg g$, the weak-localization correction $\delta G = \langle G\rangle(\beta=2) - \langle G \rangle (\beta=1)$ and the conductance fluctuations are given by\cite{footnote}
\begin{mathletters}
\begin{eqnarray} \label{eq:Gansatz}
  \delta G &=& a_1 g/\gamma + {\cal O}(g/\gamma)^{2}, \\
  \mbox{var}\, G &=& b_1 g/\gamma + b_2 (g/\gamma)^{2} + {\cal O}(g/\gamma)^{3},
\end{eqnarray}
\end{mathletters}%
where $a_1$, $b_1$, and $b_2$ are numerical coefficients determined by Eqs.\ (\ref{eq:PGavgvar}), (\ref{eq:tunnelGauss}), (\ref{eq:GINEL}), and (\ref{eq:GtunnelMom}). For the special case of two single-mode point contacts, we have
\begin{mathletters}
\begin{eqnarray}
  a_1 &=& {4 \Gamma_1^2 \Gamma_2^2 \over (\Gamma_1 + \Gamma_2)^4}, \\
  b_1 &=& {4 \Gamma_1^2 \Gamma_2^2 (\Gamma_1^2 + \Gamma_2^2 - \Gamma_1 \Gamma_2^2 - \Gamma_1^2 \Gamma_2) \over \beta (\Gamma_1 + \Gamma_2)^4}.
\end{eqnarray}
\end{mathletters}%
The coefficient $b_2$ is only relevant if $\Gamma_1, \Gamma_2 \approx 1$, when $b_1 \approx (2 - \Gamma_1 - \Gamma_2)/4 \beta \ll 1$ and $b_2 \approx (1 + 2 \delta_{\beta 1})/16$. 
At finite temperatures, in addition to dephasing, the effect of thermal smearing becomes important.\cite{Efetov1995}  Since thermal smearing has no effect on the average conductance, the weak-localization correction $\delta G$ provides an unambiguous way to find the dephasing rate $\gamma$.

The fact that dephasing was not entirely uniform in phase space in the model of Refs.\ \onlinecite{BarangerMello1995} and \onlinecite{BrouwerBeenakker1995} leads to small but noticeable differences with the completely uniform description used here, in particular for the case of single-mode point contacts. The differences may result in a discrepancy $\Delta \gamma \approx 1$ in the estimated value of the dimensionless dephasing rate $\gamma$, if the ideal voltage-probe model of Refs.\ \onlinecite{BarangerMello1995} and \onlinecite{BrouwerBeenakker1995} is used instead of the model presented here. A difference $\Delta \gamma \approx 1$ is relevant, as experiments on semiconductor quantum dots can have dephasing rates as low as $\gamma \approx 2$.\cite{MarcusPriv}

Both the voltage-probe model and the imaginary potential model only provide an effective description of dephasing. They cannot compete with a microscopic theory of inelastic scattering in quantum dots (see e.g.\ Refs.\ \onlinecite{Sivan} and \onlinecite{AGKL}). At this time, a microscopic theory for the effect of inelastic scattering on the conductance distribution does not yet exist. For the time being, the model presented here may well be the most realistic description available.

\acknowledgments

We have benefitted from discussions with I. V. Lerner, C. M. Marcus, and T.\ Sh.\ Misirpashaev. This work was supported by the ``Stich\-ting voor Fun\-da\-men\-teel On\-der\-zoek der Ma\-te\-rie'' (FOM) and by the ``Ne\-der\-land\-se or\-ga\-ni\-sa\-tie voor We\-ten\-schap\-pe\-lijk On\-der\-zoek'' (NWO).

\appendix
\section*{Calculation of {\boldmath $P(T_1,T_2)$}}

We start the calculation of $P(T_1,T_2)$ from the integral expression (\ref{eq:PS}), in which we may replace the double integral of $v$ and $v'$ by a single integral of the matrix $v' v$ over the unitary group (for $\beta=2$) or over the manifold of unitary symmetric matrices (for $\beta=1$). We make a substitution of variables $v'v \to w$ via
\begin{equation}
  v'v = \tau - \sqrt{1-\tau^2}\, w (1 - \tau w)^{-1}\, \sqrt{1-\tau^2}.
\end{equation}
The matrix $\tau$ was defined in Eq.\ (\ref{eq:PSb}). One verifies that the matrix $w$ is unitary (unitary symmetric for $\beta=1$). The Jacobian of this transformation is\cite{Brouwer1995,Hua,MPS}
\begin{equation}
  \det\left({d v'v \over d w}\right) = {V \over V'} {|\det(1 - v'v \tau)|^{\beta N_{\phi} + 2 - \beta} \over \det(1-\tau^2)^{(\beta N_{\phi} + 2 - \beta)/2}}, \label{eq:Jac}
\end{equation}
where $V$ and $V'$ are normalization constants. This change of variables is a key step in the calculation, since the Jacobian (\ref{eq:Jac}) cancels the denominator of the integrand of Eq.\ (\ref{eq:PSa}) almost completely,
\begin{eqnarray}
  P(T_1,T_2) &=& {1 \over V'} \int dw\, \Gamma_{\phi}^{\beta (6-\beta)} |T_1-T_2|^{\beta} \nonumber \\ && \mbox{} \times
  \prod_{j=1,2} (1 + \Gamma_{\phi} T_j^{-1} - \Gamma_{\phi})^{-(\beta N_{\phi} + 6 - \beta)/2} 
 \nonumber \\ && \mbox{} \times \prod_{j=1,2}
T_j^{-2\beta -2}\, |\det(1 - \tau w)|^{2 \beta}.\! \label{eq:int0}
\end{eqnarray}

We now consider separately the integral
\begin{eqnarray}
  I_{\beta} &=& \int dw\, |\det(1 - \tau w)|^{2 \beta} \nonumber \\ &=&
  \int dw\, \det (1 - \sqrt{\tau} w \sqrt{\tau})^{\beta} \det (1 - \sqrt{\tau} w^{-1} \sqrt{\tau}). \nonumber \\
\end{eqnarray}
Here we have used that $\tau$ is a positive diagonal matrix. We now change variables $\sqrt{\tau} w^{-1} \sqrt{\tau} \to \tilde w^{-1}$. If the matrix $\tau$ were unitary, we could write
\begin{eqnarray} \label{eq:int1}
  I_{\beta} &=& \int d\tilde w\, \det(1-\tau \tilde w \tau)^{\beta} \det (1 - \tilde w^{-1})^{\beta},
\end{eqnarray}
in view of the invariance of the measure $dw = d\tilde w$. However, $\tau$ is not unitary. A theorem due to Weyl allows us to continue Eq.\ (\ref{eq:int1}) analytically to arbitrary $\tau$.\cite{Weyl}

To evaluate $I_{\beta}$, we decompose $\tilde w$ in eigenvectors and eigenphases, $\tilde w = U e^{i \Theta} U^{\dagger}$, where $U$ is an orthogonal (unitary) matrix for $\beta=1$ ($2$), and $\Theta_{ij} = \delta_{ij} \theta_j$, $0 \le \theta_{j} < 2 \pi$. The invariant measure $d\tilde w$ reads\cite{Mehta}
\begin{equation}
  d\tilde w = dU \prod_{i<j} |e^{i \theta_i} - e^{i \theta_j}|^{\beta} \prod_i d\theta_i. 
\end{equation}
After some algebraic manipulations, we arrive at
\begin{mathletters} \label{eq:Iexpr}
\begin{eqnarray}
  I_{\beta} &=& \int d\theta_{1}\ldots \int d{\theta_{N_{\phi}}} 
  \prod_{i<j} |e^{i \theta_i} - e^{i \theta_j}|^{\beta} \prod_{j=1}^{N_{\phi}} \left( 1 - e^{-i \theta_i}\right) ^{\beta}\nonumber \\ && \mbox{} \times \prod_{j=1}^{N_{\phi}} \left[1 - (1-\Gamma_{\phi})e^{i \theta_j}\right]^{\beta}\, \int dU \det A^{\beta},
\end{eqnarray}
where the $2 \times 2$ matrix $A$ is given by
\begin{eqnarray}
  A_{ij} &=& \delta_{ij} - (1 - \Gamma_{\phi}) \sum_{l=1}^{N_{\phi}} {U_{il}^{\vphantom{*}} U_{jl}^{*} e^{i \theta_l}  \sqrt{T_{i} T_{j}} \over 1 - (1-\Gamma_{\phi}) e^{i \theta_l}}.
\end{eqnarray}
\end{mathletters}%

The determinant of $A$ is computed by a direct expansion. Since $N_{\phi} \gg 1$, we may consider the matrix elements $U_{kl}$ as independent real (complex) Gaussian distributed variables with zero mean and variance $1/N_{\phi}$ for $\beta=1$ ($2$). We write the result of the Gaussian integrations in terms of derivatives of a generating function $F_{\beta}$,
\begin{eqnarray} \label{eq:DiffF}
  \prod_{j=1}^{N_{\phi}} \left[1 - (1-\Gamma_{\phi})e^{i \theta_j}\right]^{\beta} \int dU \det A^{\beta} = D_{\beta} F_{\beta}.
\end{eqnarray}
The generating function $F_{\beta}$ depends on the variables $x_k$, $y_k$, and $z_k$, where $k=1$ for $\beta=1$ and $k=1,2$ for $\beta=2$,
\begin{mathletters}
\begin{eqnarray}
  && F_\beta = \prod_{j=1}^{N_{\phi}} \prod_{k=1}^{\beta} (1+x_k+y_k) [1+f(x_k,y_k,z_k) e^{i \theta_j}], ~~~~ \\
  && f(x,y,z) = (1+x+y)^{-1} (1-\Gamma_{\phi}) \nonumber \\ && \ \ \mbox{} \times \left[1 + x (1-2 T_1) + y (1-2 T_2) + z \sqrt{T_1 T_2} \right].
\end{eqnarray}
\end{mathletters}
The differential operator $D_{\beta}$ reads
\begin{mathletters} \label{eq:diffop}
\begin{eqnarray}
  D_{1} &=& \case{1}{2} N_{\phi}^{-1} (\partial_{x_1} + \partial_{y_1}) + N_{\phi}^{-2} \partial_{z_1} \partial_{z_1}, \\
 D_{2} &=& N_{\phi}^{-2} \left[ \case{1}{2} (\partial_{x_1} \partial_{x_2} + \partial_{y_1} \partial_{y_2}) - \case{1}{4} (\partial_{x_1} - \partial_{y_2})^2 \right] \nonumber \\ && \mbox{} + N_{\phi}^{-3} (\case{3}{2} \partial_{z_2} \partial_{z_2} - \case{1}{2} \partial_{z_1} \partial_{z_1})(\partial_{x_1} + \partial_{y_1}) \nonumber \\ && \mbox{} + N_{\phi}^{-4} \partial_{z_1} \partial_{z_2} \partial_{z_2} (3 \partial_{z_2} - 2 \partial_{z_1}).
\end{eqnarray}
\end{mathletters}%
The derivatives in Eq.\ (\ref{eq:DiffF}) should be evaluated at $x_k=y_k=z_k=0$ ($k=1,2$).

We are left with an integral over the phases $\theta_j$ which is of the type
\begin{eqnarray} \label{eq:II}
  I'_{\beta} &=& \int d{\theta_1} \ldots \int d{\theta_n}
  \prod_{i<j} |e^{i \theta_i} - e^{i \theta_j}|^{\beta}\nonumber \\ && \mbox{} \times \prod_{j=1}^{n} (1 - e^{-i \theta_j})^{\beta} \prod_{k=1}^{\beta} \left(a_{k} - e^{i \theta_j}\right) .
\end{eqnarray}
The integrand is a product of secular determinants $\det (\lambda - U)$ of a unitary matrix $U$. Integrals of this form were considered by Haake et al.\cite{Haake} For $\beta=1$ we can directly apply the results in their paper, for $\beta=2$ we need to extend their method to include a product of four secular determinants. We find
\begin{mathletters} \label{eq:Iresult}
\begin{eqnarray}
  I'_1 &=& {(1+n)(a_1^{n + 3}-1) - (3+n) a_1 (a^{n+1}_1 - 1) \over (a_1-1)^3 (n-1)}, \\
  I'_2 &=& {(a_1^{n + 2} - 1)(a_2^{n + 2} - 1) \over (a_1 - 1)^2 (a_2-1)^2} 
  \nonumber \\ && \mbox{} - {(a_1^{n + 2} - a_2^{n + 2})(n+2) \over (a_1-1)(a_2-2)(a_1-a_2)}.
\end{eqnarray}
\end{mathletters}%
The desired integral $I_{\beta}$ is obtained from $I'_{\beta}$ by substitution of Eq.\ (\ref{eq:Iresult}) with $n = N_{\phi}$, $a_k = f(x_k,y_k,z_k)$ into Eqs.\ (\ref{eq:Iexpr})--(\ref{eq:diffop}). Substitution of $I_{\beta}$ into Eq.\ (\ref{eq:int0}) then leads to the final result (\ref{eq:PT}).


\begin{thebibliography}{32}
\bibitem{footnote} The condition $\gamma \gg 1$ for very strong dephasing is understood as the regime $\tau_{\rm dwell} \gg \tau_{\phi} \gg \tau_{\rm erg}$, where $\tau_{\rm dwell}$ is the dwell time of the electrons in the quantum dot and $\tau_{\rm erg}$ the time scale for ergodic exploration of the complete phase space. The regime $\tau_{\phi} < \tau_{\rm erg}$ is considered in Ref.\ \onlinecite{McCannLerner}.
\end{thebibliography}
\end{document}